\documentclass[a4paper,11pt]{article}
\pdfoutput=1 
\usepackage{jinstpub} 
\usepackage{subfigure}

\title{\boldmath Production and Properties of the Liquid Scintillators used in the \textsc{Stereo} Reactor Neutrino Experiment}

\author{C. Buck,\note{Corresponding author.}}
\author{B. Gramlich,}
\author{M. Lindner,}
\author{C. Roca}
\author{and S. Schoppmann}

\affiliation{Max-Planck-Institut f\"ur Kernphysik,\\
  Saupfercheckweg 1, 69117 Heidelberg, Germany}

\emailAdd{Christian.buck@mpi-hd.mpg.de}

\abstract{The electron antineutrino spectrum in the \textsc{Stereo} reactor experiment (ILL Grenoble) is measured via the inverse beta decay signals in an organic liquid scintillator. The six target cells of the \textsc{Stereo} detector are filled with about 1800 litres of Gd-loaded liquid scintillator optimised for the requirements of the experiment. These target cells are surrounded by similar cells containing liquid scintillator without the Gd-loading. The development and characteristics of these scintillators are reported. In particular, the transparency, light production and pulse shape discrimination capabilities of the organic liquids are discussed.}

\keywords{Scintillators, Neutrino detectors}

\arxivnumber{1812.02998}

\begin{document}
\maketitle
\flushbottom

\section{Introduction}
The goal of the \textsc{Stereo} experiment is to search for an oscillation pattern at short baseline around 10~m caused by the conversion of electron antineutrinos into sterile neutrinos. The antineutrinos are detected via the inverse beta decay $\bar{\nu}_{e} + p^{+} \rightarrow e^{+} + n$ (IBD) in hydrogen nuclei (H) inside a target liquid scintillator (LS). In this reaction a positron and a neutron are produced. In the coincidence signal, the prompt positron carries the information of the neutrino energy and the delayed neutron capture emits characteristic gammas. In more than 99\% of the cases neutrons inside the target liquid are captured either by H or by gadolinium (Gd). The rare earth element Gd, which is loaded to the target LS, has the highest cross section for thermal neutron captures of all stable isotopes. Therefore, the Gd-loading significantly reduces the coincidence time. Moreover, the energy of the emitted gammas after neutron capture increases from 2.2~MeV for H to about 8~MeV for Gd. The shorter coincidence time and the high delayed energy both promote efficient background suppression in such experiments. The target LS in the \textsc{Stereo} detector~\cite{STEREO-Detector} is surrounded by a similar amount of unloaded LS to detect gammas of the neutrino signal which escape the target volume.   

The LS composition is driven by the requirements of the \textsc{Stereo} liquids. As in most multi-ton scale neutrino detectors, the organic LS solvent needs high transparency, light yield (LY) and radiopurity. In addition, it has to be compatible with the detector materials (mainly acrylic) and meet the safety aspects when operating close to a nuclear reactor (e.g.~high flash point). Those specifications led to the use of linear alkyl benzene (LAB) as the main scintillator component ($\sim$75\%). The main advantages of LAB compared to other scintillator solvents are the high transparency above 400~nm wavelength, high flash point ($\sim$140$^\circ$C), low cost and that it is chemically rather inert. However, LAB based LS have poor pulse shape discrimination (PSD) capabilities and slightly lower light yields compared to standard aromatic LS solvents. Therefore, we added $\sim$20\% of ortho-phenylxylylethane (PXE)~\cite{PXE} and 5\% of diisopropylnaphtalene (DIN) to the LS mixture. In this way, the LY and the PSD performance are improved at the cost of slightly lower transparency. 

The Gd-loading is based on $\beta$-diketone chemistry, which was developed in this context within the Double Chooz experiment~\cite{DC1}. As compared to the Double Chooz target LS~\cite{LSP}, the Gd-concentration in \textsc{Stereo} has almost doubled. This reduces the coincidence time, which is important to lower the rather high accidental background rate in the ILL environment. Such an increase in the Gd-concentration has only a minor impact on the attenuation length, but quenches the LY, as discussed in section 3. As wavelength shifting components (fluors), the quite common combination of the high quantum yield fluorescent molecules diphenyloxazole (PPO) and bis-methylstyrylbenzene (bis-MSB) is utilised. 

In the next section, we describe the LS production and the purification of the single components. Section 3 deals with specific LS characteristics, in particular the optical properties and their time stability are discussed. In section 4, the PSD properties of the liquids are reported.   

\section{Scintillator Production}
The main LS solvent component LAB could be used as delivered by Helm AG company. For the PXE (Dixie company) and DIN (Ruetasolv DI-S, R\"utgers) a dedicated column purification was performed at our institute using Al$_2$O$_3$. In particular for the PXE this purification step improved the solvent transparency significantly. The primary fluor PPO was received from Perkin Elmer (``neutrino grade'' quality). The secondary fluor bis-MSB could also be used as delivered without extra purification. 

In the first step of the target LS production, the Gd-$\beta$-diketonate compound (Gd(thd)$_3$) was mixed with tetrahydrofuran (Rotisolv, Carl Roth) at a ratio of 1:1 in weight. The Gd powder, which was purified by sublimation (Sensient company), dissolves completely in the tetrahydrofuran (THF). In parallel, the other components were mixed by mechanical stirring in a 600~litre PVDF tank in nitrogen atmosphere. First, the three solvents were blended and, then, the fluor components added as a powder. After the fluors were dissolved the Gd/THF-concentrate was poured into the tank, as well. The final LS was then filtered (0.5~$\mu$m UPE membrane) and pumped into 200~litre barrels with an inner polyethylene coating for transport to the \textsc{Stereo} experiment at ILL Grenoble. Both LS, Gd-loaded and unloaded, were produced in four batches to obtain a total amount of about 2~tons for each of them. 

\begin{table}[h]
\caption[Composition]{Composition of the \textsc{Stereo} target scintillator. The fractional concentrations are given in weight percent. The 0.9~wt.\% of Gd compound corresponds to 0.2~wt.\% of pure Gd.}
\label{Comp}
\begin{center}
\begin{tabular}{llr}
Component & Concentration & CAS Number\\
\hline
LAB (linear alkyl benzene) & 73.0~wt.\% & 67774-74-7\\
PXE (ortho-phenylxylylethane) & 19.5~wt.\% & 6196-95-8\\
DIN (diisopropylnaphtalene) & \enspace4.9~wt.\% & 38640-62-9\\
Gd(thd)$_3$	(Gd-tris-(2,2,6,6-tetramethyl- & & \\
heptane-3,5-dionate) & \enspace0.9~wt.\% & 14768-15-1\\
tetrahydrofuran (THF) & \enspace0.9~wt.\% & 1099-99-9\\
PPO (2,5-diphenyloxazole) & \enspace0.8~wt.\% \quad(7~g/l) & 92-71-7\\
bis-MSB (4-bis-(2-methylstyryl)benzene)	& \enspace0.002~wt.\% (20~mg/l) & 13280-61-0\\ 
\hline
\end{tabular}
\end{center}
\end{table}

All components of the target LS are listed in Table~\ref{Comp}. The densities were measured to be 0.887~kg/l (target LS) and 0.884~kg/l (unloaded LS) at 20$^\circ$C. For the modelling of the reflections and the light propagation in the \textsc{Stereo} detector, the refractive index, $n$, of the liquids needs to be known. An identical value of $n=1.50$ was found at 25$^\circ$C for both LS. Such a refractive index is very similar to the one of the acrylic material of the detector cells, assuring efficient light propagation to the photo sensors in the detector. 

The H fraction, an important input value to estimate the target proton number in the experiment, was measured to be $(11.45\pm0.11)$~wt.\% using CHN analysis method at TU M\"unchen. The flash point of the target liquid was determined by the Physikalisch-Technische Bundesanstalt at 74$^\circ$C. The unloaded LS for \textsc{Stereo} is identical to the target, except for the fact that the Gd-compound and the THF are missing. Without the THF the flash point is significantly higher than for the target LS. Moreover, with a PPO concentration of 3~g/l the primary fluor content is lower for the unloaded LS for reasons which are explained in the next section. 

\section{Optical Properties}
The \textsc{Stereo} experiment searches for distortions in the positron energy spectrum of the prompt neutrino signal due to oscillation effects. Since the spectral shape has to be studied with high precision, a good energy resolution is required and the energy scale needs to be known accurately. To achieve these goals, the transparency and light production of the LS are one of the main prerequisites.     

\subsection{Transparency}
The light emission of the \textsc{Stereo} scintillators is mainly in the region from $410-450$~nm. This scintillation light is observed by photosensors (8-inch Hamamatsu R5912-100 PMTs) at the top of the detector. The attenuation length in the 430~nm region should be above 4~m. Otherwise the detector resolution is degraded by vertical inhomogeneities inside the detector cells, due to significant absorption from the light produced at the detector bottom. The six components contributing to the absorbance above 400~nm, LAB, PXE, DIN, PPO, bis-MSB and the Gd-complex, all needed to be tested for optical purity before usage. The molar extinction coefficient for each of these components was determined using a UV/Vis spectrophotometer. In this way, it was possible to calculate the absorbance of the full mixture from the known concentrations of each component applying Lambert-Beer's Law~\cite{IUPAC-Gold-Book}. In principle, the proportionality of the absorbance and the concentration is only valid for low concentrations. Since the main absorbance of our LS components is well below 400~nm, the extinction at the wavelength region of interest is dominated by impurities in the chemicals. The concentrations of those impurities are very small. Therefore, we can assume the linearity between concentration and absorbance in the wavelength region of LS emission and add up the individual contributions of the single components to get the final attenuation length, as will be shown below.      

\begin{table}[h]
\caption[attenuation]{Attenuation length at 430~nm for the single components scaled to the target LS concentrations under the assumption that all the other components do not contribute to the absorption.}
\label{attMC}
\begin{center}
\begin{tabular}{l|cccccc|c}
Component & LAB & PXE & DIN & PPO & bis-MSB & Gd(thd)$_3$ & total\\
\hline
$\Lambda_{Gd-loaded}$ (m) & 23 & 35 & 33 & 41 & 176 & 94 & 7.0 \\
$\Lambda_{unloaded}$ (m) & 23 & 35 & 33 & 95 & 176 & & 8.4 \\  
\hline
\end{tabular}
\end{center}
\end{table}

In Table~\ref{attMC} the individual contributions to the attenuation length $\Lambda$ are shown. For each component with a molecular weight $M$ an attenuation length is calculated from the molar extinction coefficient $\epsilon$ and the concentration $c$

\begin{equation}
\Lambda_i = \log_{10}(e)\cdot \frac{M_i}{\epsilon_i \cdot c_i}  = 0.4343\cdot \frac{M_i}{\epsilon_i \cdot c_i}
\end{equation}

For all other components the absorbance is set to zero for comparison of the single contributions of each chemical. The total absorbance A$_{LS}$ of the liquid can be calculated by adding up the individual contributions $A_i$. Since $A\propto 1/\Lambda$, the reciprocal of the total attenuation length for the full mixture is calculated from the sum of the reciprocals of the ingredients

\begin{equation}
\frac{1}{\Lambda_{target}}=\frac{1}{\Lambda_{LAB}}+\frac{1}{\Lambda_{PXE}}+\frac{1}{\Lambda_{DIN}}+\frac{1}{\Lambda_{PPO}}+\frac{1}{\Lambda_{bisMSB}}+\frac{1}{\Lambda_{Gd}}
\end{equation}

\begin{figure}
\begin{center}
\includegraphics[width=0.6\textwidth]{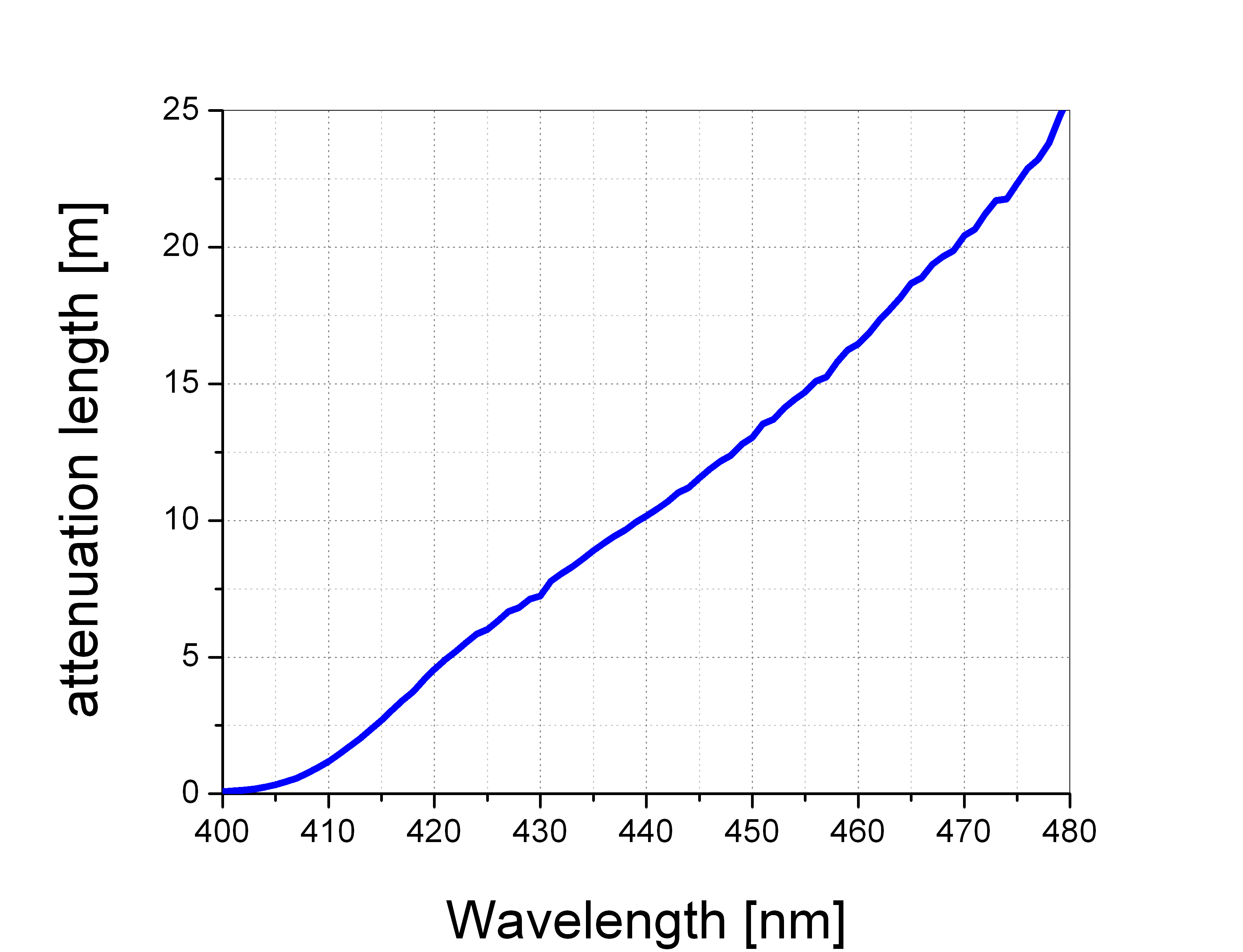}
\end{center}
\caption{\label{Fig1}Wavelength dependence of measured attenuation length for the \textsc{Stereo} target LS.}
\end{figure}

In this way, we obtain from the measured molar extinction coefficients a calculated target attenuation length of 7.0~m. This value is in very good agreement with the measured attenuation length of 6.9~m after target LS production (average value of four measurements, one for each production batch). For the case of the unloaded LS, the calculated value is 8.4~m to be compared to the measured value of 9.7~m, which is also in reasonable agreement, within the uncertainties of the measurement. The better attenuation length without Gd-loading is mainly because the PPO-concentration is more than a factor of two lower. The wavelength dependence of the target attenuation length is shown in Figure~\ref{Fig1}. Below 420~nm, self-absorption on bis-MSB starts to be dominant. In this case, the absorbed light is re-emitted with high probability. 

Since the UV/Vis measurements were done in cells up to 10~cm in length, there are limitations concerning the precision, in particular for the very transparent components which are used in high concentrations. The absorbance of the LAB at 430~nm was hardly detectable and the uncertainty in this number is, therefore, quite large. Nevertheless, the good agreements of calculated and measured values demonstrate that the presented method works and can be applied to predict the attenuation length for varying concentrations. To achieve the calculated transparency it is mandatory not to pick up any impurities from the containers or modules used in the production process. Therefore, proper cleaning and high purity of the systems involved is crucial.

For the measurement of the attenuation length in the UV/Vis spectrophotometer, scattering cannot be distinguished from pure absorption. Typical values for the scattering length of LAB based LS are in the 25~m range at 430~nm~\cite{Wurm:2010ad}. The dominant part is Rayleigh scattering for which the cross-section $\sigma$ depends on the wavelength $\lambda$, as $\sigma \propto 1/\lambda^4$. If we assumed a scattering length around 25~m also for our case, it would imply that the absorption length of the \textsc{Stereo} target liquid at 430~nm could be even up to the 10~m range.

\subsection{Light Yields}
\label{subsec:LY}
The amount of scintillation photons produced for a specific energy deposition, the LS light yield, is lowered with the addition of the Gd-complex. This complex has absorption bands in a similar wavelength region as the primary fluor and, therefore, competes with the energy transfer processes of the excited solvent molecules. Since the Gd-complex is a non-fluorescent component, energy transferred to this molecule is lost to scintillation. Therefore, the metal loading has a quenching effect on the light yield. The higher the Gd-to-primary-fluor ratio, the stronger this quenching~\cite{Ent}. 

The Gd-concentration of 0.2~wt.\% in the target LS was optimised with respect to the coincidence time of the neutrino signal. The mean capture time for neutrons in this liquid of less than 20~$\mu$s allows for an effective background reduction. The fraction of neutrons captured by the Gd isotopes is close to 90\% for such a loading. Increasing the amount of primary fluor molecules helps to reduce the quenching. Therefore, the PPO concentration was chosen to be higher for the Gd-LS (7~g/l) than for the unloaded LS (3~g/l). Above a certain concentration, self-quenching of the fluor becomes non-negligible and the gain does not increase further~\cite{Ent}. The plateau of highest light yields for the case of the \textsc{Stereo} target LS is at PPO-concentrations of about $7-10$~g/l.  

\begin{figure}
\begin{center}
\includegraphics[width=0.6\textwidth]{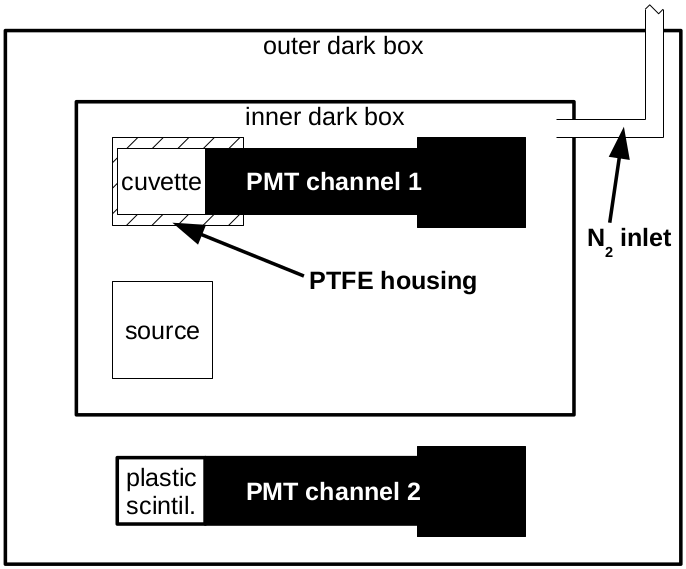}
\end{center}
\caption{\label{DarkBox}Sketch of the setup to measure light yield and pulse shape discrimination properties. PMT channel 2 is only used in the backscatter configuration.}
\end{figure}

The light yield of the \textsc{Stereo} liquids was measured inside a dark box by irradiating LS samples in a 86~ml cuvette made from Pyrex glass (Starna 35-PX-50) with a ${}^{60}$Co gamma source of $\sim$300~kBq activity. Resulting LS pulses were detected by a 2-inch PMT (ET Enterprises 9954B) covering the entire base surface of the cylindrical cuvette (see Figure \ref{DarkBox}). The utilised PMT features good timing as well as high quantum efficiency between 320 and 500~nm, matching the emission spectrum of the LS. Optical coupling between PMT and cuvette was achieved through optical grease. Cuvette and PMT cathode were fitted into a PTFE housing allowing for high reflectivity. To prevent contamination by external quenchers like oxygen, all LS samples were purged with nitrogen before sealing and subsequently kept under nitrogen atmosphere at 19$^\circ$C during the measurement. For the measurement, a backscatter setup was used by demanding a coincident signal in a plastic scintillator block located opposite to the LS sample (see Figure \ref{DarkBox}). With this technique, mainly those events are accepted, where a gamma from the ${}^{60}$Co source first scatters off an electron of the LS sample (channel 1) at an angle of 180${}^{\circ}$ and then deposits energy in the plastic scintillator (channel 2). Gammas having other scattering angles deposit less energy in the LS sample, but they are eventually missing the scintillator block and are, thus, rejected. The resulting charge distribution measured in channel 1 shows a clear peak corresponding to the known Compton-backscatter energy of ${}^{60}$Co, allowing for energy calibration and relative light yield comparison of different LS samples.

In the measurements, the unloaded LS shows a factor of $(1.30\pm0.12)$ higher light yield compared to the Gd-loaded LS. From an independent MC simulation of the \textsc{Stereo} detector, the ratio of light yields between unloaded and Gd-loaded LS is 1.30, which is in perfect agreement with the experimental value. Relative measurements were also performed compared to a pure LAB-based LS with 3 g/l PPO and 20 mg/l bis-MSB as secondary wavelength shifter. A light yield of about 50\% anthracene is reported for this liquid~\cite{DBLS}. Anthracene itself produces about 17400 photons/MeV. Using these numbers we obtain absolute light yield values of 6500 photons/MeV for the Gd-loaded LS and 8400 photons/MeV for the unloaded LS.

Investigating the effect of single LS components on the light yield (see Table \ref{tab:PSD}) we find that the addition of few percent DIN increases the light yield substantially for the Gd-loaded LS while leaving the light yield of the unloaded LS unchanged. This could be explained by the light absorption spectrum of DIN around 280~nm \cite{QuantumYields}. Together with PPO, it concurs for absorption with the quenching molecule Gd(thd)$_3$. Since DIN is a solvent, its concentration can be chosen to be substantially higher than the PPO concentration (cf. Table \ref{Comp}). Thus, DIN allows for a stronger mitigation of quenching by Gd than PPO alone.

\subsection{Stability}
The stability of the optical properties, in particular of the transparency, is one of the main challenges for metal loaded liquid scintillators. Liquid scintillators based on the $\beta$-diketone chemistry, as the \textsc{Stereo} target liquid, are known for their stability and purity~\cite{LSP}. Typically, the attenuation length is more affected by degradation over time than the light yield. The attenuation lengths of the \textsc{Stereo} liquids were first measured in a UV/Vis spectrophotometer after production in July/August 2016. Then, they were re-measured in April 2017. Samples were taken directly from the detector when it had to be emptied for reactor maintenance. For the target LS the measurement in April 2017 resulted in a value of $(6.1\pm0.5)$~m to be compared to the $(6.9\pm0.6)$~m measured about 7 months before. For the unloaded scintillator it was found to be $(9.9\pm1.1)$~m, whereas we had $(9.7\pm1.1)$~m before detector filling. 

Those numbers demonstrate the stability and robustness of the \textsc{Stereo} liquids. No significant degradation is observed during the investigated time period of more than half a year including critical liquid handling operations, as the liquid transport in barrels from MPIK in Heidelberg to ILL in Grenoble, as well as filling and emptying of the \textsc{Stereo} detector. The liquid temperature during data taking between November 2016 and March 2017 was around 25$^\circ$C. Stability checks with test samples stored in the laboratory might not be representative, since the storage conditions and the handling are different from those of the detector liquids. If degradation is observed it is often not due to intrinsic instabilities under ideal conditions, but related to environmental effects or LS processing.        

\section{Pulse Shape Discrimination}
To discriminate correlated background in the \textsc{Stereo} experiment, time shape information of scintillation pulses is used. One critical background class is due to fast neutrons. Here, the prompt event is mimicked by recoil protons and the delayed by the neutron capture in Gd. Such background events can be distinguished from IBD events by a longer component in the time shape of the scintillation pulse. The recoil protons excite more triplet states of the solvent molecules than positrons. These triplet states have longer decay times increasing the charge ratio in the tail of the scintillation pulse. Initially, we found that admixtures of DIN at the few per cent level improve the pulse shape discrimination (PSD) capabilities of the LS~\cite{Col}. Similar observations were also made for the LS in the NEOS experiment~\cite{Kim:2015pba}. 
\begin{figure}[tb]
\begin{center}
\subfigure[]{
	\includegraphics[width=0.48\textwidth]{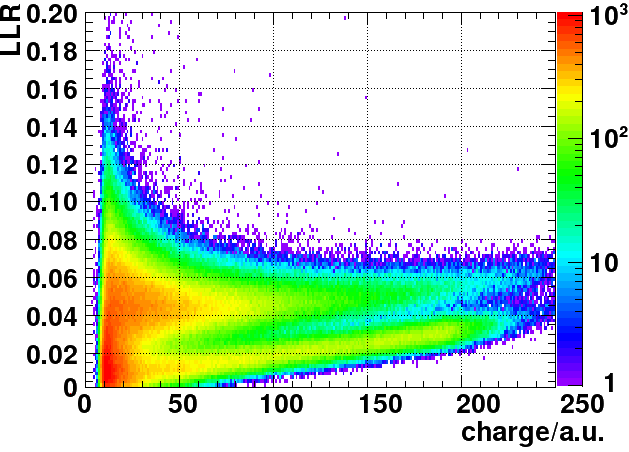}
	\label{Fig:PSD:Plane}
	}
\subfigure[]{
	\includegraphics[width=0.48\textwidth]{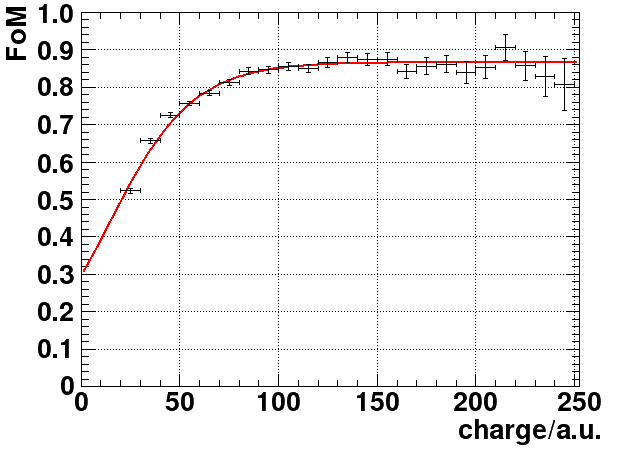}
	\label{Fig:PSD:Energy}
	}
\end{center}
\caption{\label{Fig:PSD}(a) PSD distribution in the late light ratio (LLR)-charge plane and (b) charge dependent figure of merit (FoM) for the \textsc{Stereo} target LS. In both plots, a charge of 48 units corresponds to 1~MeV gamma energy (${}^{60}$Co backscatter peak).}
\end{figure}

To measure the PSD properties of the \textsc{Stereo} scintillators the same setup as for the light yield measurement was used (cf. Section \ref{subsec:LY}). However, no backscatter coincidence was required. The LS samples were irradiated with gammas and neutrons from an AmBe source of 1.11~GBq activity (70000 neutrons/s). To quantify the PSD ability of the LS, the late light ratio (LLR) of the scintillator pulses was used. It is defined as the quotient of the charge after the first $n$ nanoseconds and the total charge of the pulse. The time-parameter $n$ was optimised to give the best separation between gamma and neutron events. The separation power is given in terms of the Figure of Merit (FoM) defined as $\textrm{FoM} = \frac{1}{2 ~ \sqrt[]{2 \cdot \ln(2)}} \frac{\vert\mu_{n} - \mu_{e}\vert}{\sigma_{n} + \sigma_{e}}$ with $\textrm{N}(\mu_{x},\sigma_{x}), ~ x \in \lbrace n, e \rbrace$ denoting two normal distributions modelling the nuclear and electronic recoil population, respectively. 

The measurements show a clear separation between the electronic recoil population at a LLR of 0.02 and the nuclear recoil population at 0.05 (see Figure \ref{Fig:PSD:Plane}). In this plot, a charge of 48 units corresponds to a gamma energy of 1 MeV (${}^{60}$Co backscatter peak). As the mean values of the LLR distributions are evolving with charge, the FoM is calculated for charge bins of 10 units width up to the end of the spectra, at about 250 charge units. A logistic function is fitted to the data (see Figure \ref{Fig:PSD:Energy}) and an asymptotic FoM of $0.865 \pm 0.055$ is achieved for the Gd-loaded LS. The achieved FoM in \textsc{Stereo} lies below this value at 0.71 \cite{STEREO-Detector} mainly due to the larger dimensions of the detector cells, resulting in broadening of light pulses.

Additional measurements for similar LS compositions were carried out and are given in Table \ref{tab:PSD}. Among the two \textsc{Stereo} LS, the Gd-loaded LS performs substantially better compared to its unloaded counterpart. PSD in the Gamma Catcher, i.e. the volume filled with unloaded LS, is not required for the \textsc{Stereo} experiment, thus the unloaded LS was not optimised in terms of PSD capabilities. Looking at individual components, PXE can be identified as reason for an improved FoM. It even overcompensates the loss in FoM due to Gd in the LS. In contrast, the combination of PXE and DIN causes a decrease of FoM for the unloaded LS, while each component individually yields an improvement. So the positive effect of DIN on the PSD as indicated in~\cite{Col} could not be confirmed by these studies. 

\begin{table}[h]
\caption[PSD]{Comparison of figures of merit (FoM) and light yield (LY) in photons/MeV for different liquid scintillators. All values have uncertainties of $\pm 6\%$ (FoM) and $\pm 5\%$ (LY), respectively. Unless stated otherwise and with exception of LAB, the concentrations of each component are those of Table \ref{Comp}. The concentration of LAB is always chosen such that the sum of concentrations yields 100\%.}
\label{tab:PSD}
\begin{center}
\begin{tabular}{rrrrrrr|r|r}
\multicolumn{7}{c|}{liquid scintillator composition}& FoM & LY\\
\hline
\multicolumn{9}{c}{laboratory measurements}\\
\hline
LAB&    &    &            &    &7g/l PPO& bis-MSB& 0.76&9000\\
LAB&    &    & Gd(thd)$_3$& THF&7g/l PPO& bis-MSB& 0.67&4300\\
LAB& PXE&    & Gd(thd)$_3$& THF&7g/l PPO& bis-MSB& 0.87&6000\\
LAB& PXE& DIN& Gd(thd)$_3$& THF&7g/l PPO& bis-MSB& 0.87&6500\\
\hline
LAB&    &    &            &    &3g/l PPO& bis-MSB& 0.76&8700\\
LAB& PXE&    &            &    &3g/l PPO& bis-MSB& 0.96&8400\\
LAB&    & DIN&            &    &3g/l PPO& bis-MSB& 0.81&7900\\
LAB& PXE& DIN&            &    &3g/l PPO& bis-MSB& 0.71&8400\\
\hline
\multicolumn{7}{l|}{100\% BC-501A}& 1.06 &\\
\hline
\multicolumn{9}{c}{\textsc{Stereo} in-situ~\cite{STEREO-Detector}}\\
\hline
LAB& PXE& DIN& Gd(thd)$_3$& THF&7g/l PPO& bis-MSB& 0.71&\\
\hline
\end{tabular}
\end{center}
\end{table}

\section{Summary} 
The \textsc{Stereo} liquid scintillator production was optimised to obtain ``safe'' liquids with high transparency, reasonable pulse shape discrimination capability and long term stability. We developed liquids with more than 6~m attenuation length above 430~nm with a Gd-loading of 0.2~wt.\% for the case of the target LS. No significant degradation could be observed, at least within the first months of data taking. We found the light yield to be improved by the admixture of few per cent of DIN. Moreover, the admixture of PXE yields a good pulse shape discrimination capability. All basic \textsc{Stereo} requirements of the liquids were met providing the basis for successful neutrino detection in a nuclear reactor~\cite{Stereo_PRL}.  

\section*{Acknowledgements}
We would like to thank Honghanh Trinh Thi, TU M\"unchen (Germany) for help on the UV/Vis measurements and the CHN analysis. We also acknowledge the help of David Lhuillier and Maxime P\'equignot from IRFU Saclay (France) for preliminary measurements and a first characterisation of \textsc{Stereo} liquid samples. Finally, we would like to thank the whole \textsc{Stereo} Collaboration for valuable discussions and support.

\end{document}